\documentclass[a4paper,fleqn]{cas-dc}

\usepackage[numbers, sort&compress]{natbib}
\usepackage{balance}
\usepackage{amsmath}
\usepackage{amssymb}
\usepackage{comment}
\usepackage{siunitx}
\usepackage{hyperref}
\usepackage{xurl}
\usepackage{url}
\usepackage{placeins}
\usepackage{graphicx}
\usepackage{placeins}
\usepackage{chngcntr}
\def\tsc#1{\csdef{#1}{\textsc{\lowercase{#1}}\xspace}}
\tsc{WGM}
\tsc{QE}

\begin{document}
\let\WriteBookmarks\relax
\def\floatpagepagefraction{1}
\def\textpagefraction{.001}

\shorttitle{}    

\shortauthors{Raisch et al.}  


\title [mode = title]{Very Exciting: Zero-Shot Model Predictive Control of Buildings via Excitation-Based Generalized Transfer Learning Models} 


%

\author[1, 2]{Fabian Raisch}
\cormark[1]
\cortext[cor1]{Corresponding author}

\ead{fabian.raisch@th-rosenheim.de}

\credit{Conceptualization, Software, Methodology, Investigation, Data curation, Validation, Visualization, Writing – Original Draft, Review \& Editing}
\affiliation[1]{organization={Technical University of Applied Sciences Rosenheim}, city={Hochschulstraße 1, Rosenheim, 83024}, country={Germany}}
\affiliation[2]{organization={Technical University of Munich}, city={Arcisstraße 21, Munich, 80333}, country={Germany}}

\author[1, 3]{Felix Koch}
\credit{Conceptualization, Software, Methodology, Writing – Review \& Editing}
\affiliation[3]{organization={Karlsruhe Institute of Technology}, city={Hermann-von-Helmholtz-Platz 1, Eggenstein-Leopoldshafen, 76344}, country={Germany}}
\author[4]{Zack Xuereb Conti}\credit{Conceptualization, Writing – Review \& Editing}
\affiliation[4]{organization={Alan Turing Institute}, city={96 Euston Rd., London, NW1 2DB}, country={UK}}
\author[2]{Christoph Goebel}\credit{Supervision, Writing – Review \& Editing}
\author[1]{Benjamin Tischler}\credit{Supervision, Conceptualization, Writing – Review \& Editing}


\begin{abstract}
The widespread adoption of data-driven, energy-efficient model predictive control (MPC) in buildings remains hindered by substantial effort to collect data and train models for individual buildings. Transfer learning (TL) has consequently gained increasing attention for target building modeling, as it reduces data requirements and modeling effort by reusing pretrained source models. However, these TL models are typically evaluated only on prediction accuracy in the target, without testing downstream control performance.
To address this gap, we apply a state-of-the-art TL approach -- pretraining a generalized model on multiple source buildings using standard operational data -- within an MPC setup in a target building. We show that this approach is insufficient to achieve satisfactory control performance. As a solution, we introduce generalized models pretrained on excitation-based operational source data -- purposefully probed inputs that explore the building's state-action space. For evaluation, we apply the generalized models via zero-shot (i.e., without fine-tuning) to 32 simulated target buildings and assess MPC performance. Our results show that excitation-based generalized models achieve the strongest control performance among all benchmarks, outperforming an online linear model-based MPC and a PI controller by 6.4\% and 36.9\%, respectively. By combining strong control performance with the ability to generalize across multiple buildings, without requiring any target-specific data, our approach reduces MPC setup cost and simplifies its widespread deployment in the building sector.

\end{abstract}





\maketitle

\section{Introduction}
\label{sec_intro}

Thermal energy consumption in households alone accounts for approximately 22\% of European energy demand \cite{eea2023DecarbonisingHeatingCooling}. Advanced control strategies, such as Model Predictive Control (MPC), can reduce this demand by 10--50\% \cite{Drgona.2020, Hilliard02072016}. 
MPC relies on a model of the building’s thermal dynamics to predict how indoor temperatures evolve over a future horizon in response to candidate control actions. Since thermal characteristics vary considerably across buildings, an individual model is typically required for each building.
Data-driven and machine learning (ML) approaches have therefore emerged as a promising path toward scaling MPC to widespread application. Unlike physics-based modeling, data-driven modeling reduces the need for domain expertise and, thus, manual engineering effort \cite{Drgona.2020}. However, this scalability comes at a recurring effort: for every new building, large amounts of data must be collected and a model trained \cite{KHABBAZI2025126459, Zhan2021}.
To address these limitations, transfer learning (TL) methods have emerged to make data-driven models transferable and reusable across multiple buildings \cite{pinto2022transfer, peirelinck2022transfer}.
TL methods use knowledge from a source building as a starting point to model an unknown target building. As a result, the data collection process in the target can be reduced or omitted, no new model needs to be trained, and prediction performance can be improved compared to training a model from scratch \cite{RAISCH2026116868, li_building_2024, pinto_sharing_2022}. In particular, pretraining a model on data from multiple source buildings -- a \textbf{generalized model} -- has been shown to achieve superior prediction performance across a large variety of target buildings compared to relying on a single source building or on time series foundation models (TSFMs) \cite{raisch2025gentlgeneraltransferlearning, koch2026thermalgems, fm_nagy}. 
While these TL studies are primarily motivated by enabling downstream MPC, they are evaluated solely on target prediction performance, without assessing the resulting control performance. This fact is therefore critical, since good prediction accuracy does not necessarily lead to good control performance (prediction--control performance gap) \cite{BLUM2019410, stoffel_real-life_2024} -- a phenomenon also introduced as the ``accuracy paradox'' \cite{martin_poster}.
This phenomenon motivates the following research question: \textit{To what extent are generalized models suitable for downstream control applications?}

Beyond TL-based modeling, numerous studies have demonstrated the benefits of training models on excitation-based data, as opposed to training models on standard operational data \cite{Drgona.2020, SERASINGHE2024114123}. Standard operational data result from set-point-driven building operation using standard controllers, such as hysteresis and PI controllers. As a result, only a narrow region of the state-action space near the temperature setpoint is covered. Training models on this kind of data can lead to inaccurate thermal predictions in unseen operating conditions \cite{koch2026buildynexcitationdrivendatageneration}. Excitation-based approaches address this limited state-action space by probing the building with purposefully designed actuation signals. Using data obtained from such excitation strategies for model generation has been shown to improve both prediction and control performance compared to models trained on standard operational data \cite{bacher_identifying_2011, madsen1993short, 100c2e3b8bb24a83bab3ae769708cec8, knudsen_experimental_2021, SARTORI2023109149, koch2026buildynexcitationdrivendatageneration}. 
Nevertheless, excitation strategies can only be applied partially in occupied buildings, to avoid comfort violations and increased energy consumption, and to limit the additional time and effort required for planning \cite{SERASINGHE2024114123, Drgona.2020}.
TL, however, offers a natural solution to these limitations: since pretraining is performed on source buildings, excitations can be applied across the sources without disrupting target occupants. Hence, training a generalized model, as introduced above, but with different excitation strategies in the sources, could further improve generalization not only across different buildings but also over a larger state-action space. Once trained, the excitation-based generalized model could be applied to control an unseen target building. 
Yet, it remains unclear whether this approach improves MPC performance and thereby narrows the prediction--control performance gap identified in the accuracy paradox. This leads to the second research question: \textit{Does incorporating excitation strategies during pretraining generalized models improve control performance in target buildings?} 

Furthermore, studies investigating ML models for building control often conclude that linear models achieve superior control performance over neural networks \cite{ stoffel_real-life_2024, BUNNING2022118491}. This contrasts with the fact that buildings exhibit nonlinear dynamics \cite{Drgona.2020, BLUM2019410}, as well as with findings that neural networks provide higher predictive accuracy than linear models in buildings \cite{stoffel_real-life_2024, TSFM_zero, choi_performance_2023}. This again highlights the prediction--control performance gap: neural networks with higher predictive accuracy do not necessarily translate into better control performance. Hence, it remains unclear whether generalized models can compete with or even outperform linear models in downstream control applications. 



\subsection{Literature Review}
\label{sec_lit}

This literature review is structured around three complementary areas that are central to this work: MPC strategies for buildings, excitation strategies for system identification, and transfer learning for building thermal dynamics modeling.

\textbf{MPC strategies:} Several studies have investigated advanced MPC strategies for buildings. Most of them are based on data-driven models, as white-box (pure physical) models are not scalable for broad application \cite{Drgona.2020}. RC-gray-box models are therefore often employed, which approximate the building as a simplified resistance--capacitance network with parameters identified from data \cite{SERASINGHE2024114123}. For example, Huang et al. \cite{huang_model_2014} employed an RC model of an airport terminal building within an MPC framework and reported energy savings of 5--18\%. Nevertheless, RC models still require physical domain knowledge and face difficulties in inverse modeling, as parameter identification typically involves non-convex loss landscapes, dependence on good initial guesses, and sensitivity to training data quality \cite{SERASINGHE2024114123, raisch2026transferlearningneuralparameter, BLUM2019410}. As a result, purely black-box models have recently gained wider adoption. To systematically compare these modeling paradigms, Stoffel et al. \cite{STOFFEL2023112709} evaluated a white-box, an RC-based gray-box, and a black-box model for MPC, and additionally assessed a reinforcement learning approach. Among the evaluated methods, the black-box multi-layer perceptron (MLP) model achieved the best overall control performance, whereas RL yielded the lowest savings. 
In a follow-up study, Stoffel et al. \cite{stoffel_real-life_2024} focused specifically on the black-box category, comparing an MLP, a Gaussian process regression model, and a multi-input linear regression model. Here, the MLP achieved the highest prediction accuracy, but this did not translate into the best control performance. The linear model, by contrast, achieved the best control performance despite slightly weaker predictive accuracy and was less sensitive to input feature selection. Consequently, the authors recommended the linear model for control. 
Similarly, Bünning et al. \cite{BUNNING2022118491} recommended a linear model. They compared a linear Autoregressive--Moving-Average with Exogenous Inputs (ARMAX) model with Random Forests and Input Convex Neural Networks. Their MPC framework achieved heating and cooling energy savings between 26\% and 49\% compared with a hysteresis baseline controller. The authors recommended the linear ARMAX model due to its lower computational requirements and superior sample efficiency.

Overall, the current literature favors linear models because of their simple setup, low computational complexity, and reliable control performance. At the same time, ML-based models achieve higher prediction accuracy, suggesting potential to also improve control performance. The studies in \cite{stoffel_real-life_2024, STOFFEL2023112709, BUNNING2022118491} are limited by the evaluation of only a single neural network architecture, the use of a small number of buildings, and the absence of transfer learning approaches.

\textbf{Excitation strategies:} Pseudo Random Binary Sequence (PRBS) excitation is the most commonly used excitation strategy in buildings \cite{SERASINGHE2024114123, Drgona.2020}. PRBS is a deterministic signal with white-noise characteristics that switches the heating system between on and off states, thereby exciting a broad range of system dynamics.
For example, Royer et al. \cite{royer2014procedure} applied a PRBS signal to the temperature setpoint in an EnergyPlus simulation, as direct access to the controller was unavailable. The resulting data were used to train linear black-box models, achieving normalized prediction accuracies between 78\% and 93\%.
Madsen and Schultz \cite{madsen1993short} and Bacher and Madsen \cite{bacher_identifying_2011} proposed using two coupled PRBS schemes with different switching intervals to excite short and long time constants and capture air and envelope dynamics for better estimation results.
Beyond PRBS, Jain et al. \cite{Jain_ICCPS_excite} compared uniform random sampling and a custom Optimal Experiment Design (OED) strategy for training Gaussian Process models. While OED and uniform random sampling generally achieved better results than PRBS, the performance differences decreased with increasing excitation duration.

Overall, dedicated excitation strategies have shown promising improvements in system identification. However, existing studies are largely limited to individual buildings and relatively simple model architectures, leaving their applicability to TL and advanced ML models largely unexplored.

\textbf{Transfer learning for buildings:} Several studies have employed TL strategies for buildings to address the scalability and data-efficiency issues \cite{pinto2022transfer}. Some focus on single-building, real-world case studies \cite{DOU2025113341, cho2024application, chen2020transfer}, while others consider a larger number of target buildings, albeit simulated \cite{chaudhary2025transfer, raisch2025gentlgeneraltransferlearning, fm_nagy}. An important challenge in TL is selecting a suitable source building from which knowledge can be transferred to a target. To address this issue, Li et al. \cite{li_building_2024} investigated different source selection strategies and found that randomly selecting a source from the same building domain as the target yields the best performance. Building on this finding, Raisch et al. \cite{raisch2025gentlgeneraltransferlearning} proposed a generalized model approach in which a model is pretrained on data from multiple source buildings. The resulting generalized model outperformed both fine-tuning randomly selected source models, as done in \cite{li_building_2024}, and models trained from scratch. Extending this work, Koch et al. \cite{koch2026thermalgems} evaluated generalized models across multiple datasets, including real-world measurements, comparing them against state-of-the-art TSFMs. Depending on the dataset, 16--64 source buildings were required for the generalized model to outperform a TSFM in a target building.

Overall, these TL studies demonstrate the potential of TL for modeling building thermal dynamics. However, two gaps remain unaddressed. First, none of the existing studies investigate the role of dedicated excitation strategies during pretraining. Second, their evaluation focuses solely on predictive performance, leaving the impact of TL on downstream MPC performance unexplored. Although Kim et al. \cite{kim2024model} propose a TL-based control approach using MPC, the transfer is limited to adapting a model from simulation to a real building. Similarly, Wan et al. \cite{wanmodel} refer to TL-based MPC, but the transfer occurs only between different operating conditions of the same building, which are treated as source and target domains. Both studies, therefore, do not follow the standard TL setup considered in this work, which involves transferring knowledge between independent source and target buildings.
\subsection{Contribution}
\label{sec_gap}

We pretrain generalized models on data from 64 simulated source buildings, using different neural network architectures (MLP, LSTM, and Transformer) and either standard operational data or one of three excitation strategies (PRBS, a ramping strategy, or a random walk). We evaluate these generalized models in zero-shot MPC (i.e., without fine-tuning) on 32 unseen target buildings, benchmarking their control performance against an online linear model-based MPC and a PI controller. All methods are assessed for both control and prediction performance, the latter under two testing protocols: standard operational data and data obtained during actual MPC operation.

With this work, we bring TL models to the downstream task of MPC, aiming to advance research towards broader and faster applicability of energy-efficient control in buildings. 
The main contributions of this paper are:
\begin{itemize} 
    \item This is the first study, to the best of our knowledge, to investigate generalized TL models in a closed-loop MPC setup.
    \item With excitation-based pretraining, we enable effective zero-shot control in target buildings while narrowing the prediction--control performance gap.
    \item The excitation-based generalized model outperforms an online linear model-based MPC by an average 6.4\% reduction in MPC cost, while requiring no target-specific data.
    
\end{itemize}

The remainder of this paper is structured as follows: Section~\ref{sec_method} presents the methodology, Section~\ref{sec_results} the experiments and results, Section~\ref{sec_discuss} the discussion, and Section~\ref{sec_conclusion} the conclusion.


\section{Method}
\label{sec_method}

\begin{figure*}[t]
	\centering
	\includegraphics[width=.8\textwidth]{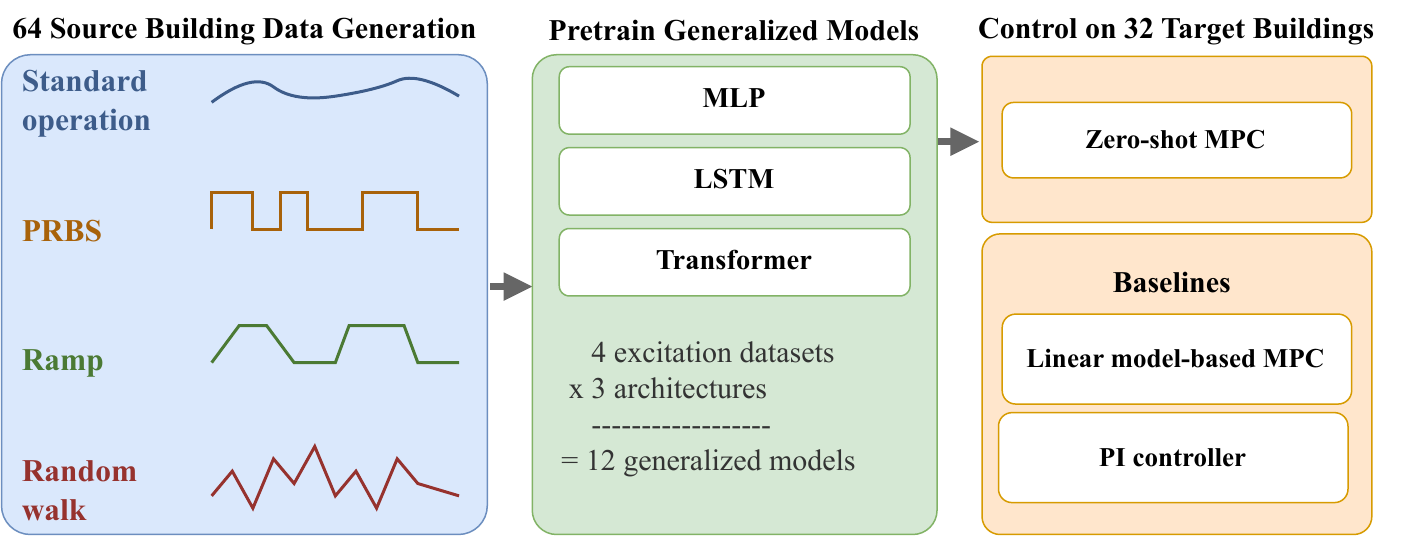}
	\caption{Methodological overview.}
	\label{fig_method}
\end{figure*}

This section describes the overall method of the paper, which follows Figure~\ref{fig_method}. Section~\ref{sec_data_gen} describes the simulation environment used for the source and target buildings. Section~\ref{sec_target_control} explains the general MPC setup applied to the target buildings. Section~\ref{sec_pretraining} demonstrates the pretraining of the generalized models and their application to target control. Finally, Section~\ref{sec_eval} defines the evaluation setup and the baseline controllers.

\subsection{Building-Simulation Environment}
\label{sec_data_gen}

Pretraining generalized models and evaluating them in closed-loop MPC requires access to a large and diverse set of buildings. To enable evaluation across numerous buildings for MPC and to enable comparable settings across control conditions, a simulation environment is employed. 
We use the BuilDa simulation environment \cite{builDaReview, builda2}, which allows adjustments for building properties, occupancy, weather, and control actions. The environment is based on a Modelica simulation and is exported as a functional mock-up unit (FMU) to enable building adjustments and simulations within Python. The BuilDa simulation has been validated against the ANSI/ASHRAE 140-2004 test cases TC600, TC900, TC600FF, and TC900FF \cite{ASHRAE140_2004} and has already been applied in \cite{RAISCH2026116868, raisch2025gentlgeneraltransferlearning, raisch2026transferlearningneuralparameter, koch2026thermalgems, koch2026buildynexcitationdrivendatageneration}.
The base building configuration consists of a single-zone model with two floors, a roof, ceiling, floor, exterior and interior walls, furniture, and windows facing each cardinal direction. The thermal-energetic properties of these elements can be individually configured. With this variability, the BuilDa simulation can represent a large population of single-family houses \cite{builda2, TASK442013}. The heating system is represented by an ideal thermal source, allowing different heating systems to be represented. Based on the building configuration, the nominal heating power (employed as maximum power) is determined from the transmission and ventilation losses through the building envelope, as in \cite{normHeizlast}. This ensures that every simulated building is equipped with an appropriately dimensioned heat source.

For this study, we use a distinct set of source buildings for pretraining and a separate set of target buildings for testing. To this end, we adopt the building configuration distribution for the sources and targets from \cite{raisch2025gentlgeneraltransferlearning, RAISCH2026116868}, who employed a similar TL approach. Table~\ref{tab_permutations} summarizes the configurations for both sets. These buildings represent the domain of single-family houses in Central Europe, built between 1949 and today \cite{tabula}, while differing in building properties and locations between sources and targets. 
For the sources, we randomly sample a subset of 64 buildings from their distribution. This number follows Koch et al. \cite{koch2026thermalgems}, who show it to be sufficient for pretraining a generalized model. For the targets, we sample 32 buildings to balance evaluation robustness and computational cost during experiments. 

\begin{table}[!b]
    \caption{Parameter distribution for the source and target buildings, adapted from \cite{raisch2025gentlgeneraltransferlearning, RAISCH2026116868}, including the insulation level of the exterior wall ($U\text{-}value_{\text{wall}}$), the area-specific heat capacity of the exterior wall ($c_{\text{wall}}$), the window-to-wall area ratio ($f_{\text{win}}$), and the building ground area ($A_{\text{ground}}$). The $U$-values for the windows and the roof are simplified to $U\text{-}value_{\text{win}} = 1 + U\text{-}value_{\text{wall}}$ and $U\text{-}value_{\text{roof}} = U\text{-}value_{\text{wall}}$, respectively. The remaining simulation parameters are calculated by BuilDa and correspond to the respective energy-efficiency levels depending on the selected values from this table \cite{builda2}.}
\label{tab_permutations}
    \centering
    \begin{tabular}{p{2cm}|p{2.5cm}|p{2.5cm}}
        \textbf{Parameter} & \textbf{Sources} & \textbf{Targets} \\ \hline
        $U-value_{\text{wall}}$ [W/(m\textsuperscript{2}K)] & $\{0.1, 0.4, 0.7, 1, 1.3\}$ & $\{0.25, 0.55, 0.85, 1.15\}$ \\ \hline
        $c_{\text{wall}}$ [kJ/(m\textsuperscript{2}K)] & $\{30, 165, 300\}$ & $\{40, 150, 280\}$ \\ \hline
        $f_{\text{win}}$ [-] & $\{0.15, 0.2\}$ & $\{0.16, 0.19\}$ \\ \hline
        $A_{\text{ground}}$ [m\textsuperscript{2}] & $\{60, 90, 120\}$ & $\{70, 100\}$ \\  \hline
        Weather & Belgrade, Prague, Berlin, London, Zurich & Munich, Amsterdam,  Bratislava \\
    \end{tabular}
\end{table}

It should be noted that, for a real-world application of the proposed method, obtaining data from 64 real source buildings would be challenging. However, the underlying idea of this paper is for the source data generation process to take place at the simulation level, while target control is prospectively intended for a real building, as further discussed in Section~\ref{sec_limits}.

\subsection{General MPC Formulation}
\label{sec_target_control}


This section explains the closed-loop MPC setup for the target buildings. We extend the BuilDa simulation environment with an MPC controller, as only standard P, PI, and hysteresis controllers were included. The code for the controller, including the experiments from Section~\ref{sec_results}, can be found at \cite{github_exciting}. For MPC, we use the following formulation:

\begin{subequations}
\label{eq_mpc}
\begin{align}
\min_{\mathbf{u}, \mathbf{\epsilon}} \quad
& \sum_{k=0}^{H_p-1} \Big[ w_0\epsilon_k^2 + w_1 \, u_k^2 + w_2(u_k - u_{k-1})^2 \Big] \label{eq_mpc_obj}\\
\text{s.t.} \quad
& T_{k+1} = f_\theta\!\left(T_{k},\, d_{k},\, u_{k}\right), \label{eq_mpc_model}\\
& T_k^{\text{low}} - \epsilon_k \leq T_k \leq T_k^{\text{up}} + \epsilon_k, \label{eq_mpc_comfort}\\
& \epsilon_k \geq 0, \label{eq_mpc_slack}\\
& 0 \le u_{k} \le 1, \label{eq_mpc_input}\\
&\forall k \in [0, \dots, H_p-1] \label{eq_mpc_k}
\end{align}
\end{subequations}

Here, $k$ denotes the discrete time step within the prediction horizon $H_p$, $f_\theta(\cdot)$ denotes the prediction model of the building dynamics, $T_k$ and $T_{k+1}$ denote the current and future indoor temperature states, $d_k$ denotes disturbances such as weather, and $u_k$ is the heating control signal. We use $u_k$ as a normalized control signal to make energy use comparable across different nominal powers of buildings; in the remainder of this paper, we therefore refer to this quantity as \emph{heating activity} rather than energy use. The objective function follows \cite{stoffel_real-life_2024, STOFFEL2023100296} and comprises three terms: a penalty on temperature band violations ($w_0$) via the slack variable $\epsilon_k$, a term for minimizing heating activity ($w_1$), and a smoothing term penalizing changes in the control signal to avoid oscillations ($w_2$). The temperature bounds are time-varying, defined by $T_k^{\text{low}}$ and $T_k^{\text{up}}$ and set according to Table~\ref{tab_Tset}, similar to the BOPTEST setup \cite{blum2021building, counterdyna}. 

Cost minimization based on electricity price is another possible objective. However, we deliberately omit it here, as it would prevent a direct comparison with \cite{stoffel_real-life_2024}, which we use as our primary benchmark. Furthermore, price signals introduce an additional degree of freedom, i.e., the choice of price signal. Also, cost signals can visually clutter results, as the resulting control actions are harder to interpret directly. 


\begin{table}[tbh]
    \caption{Comfort range and set points for the indoor temperature.}
    \label{tab_Tset}
    \centering
    \begin{tabular}{p{3.7cm}|p{1.5cm}|p{1.5cm}}
    Time & $T^{\text{low}}$  & $T^{\text{up}}$ \\ \hline
    Mon. -- Fri. 6:00--18:00 & 17 \si{\celsius} & 24 \si{\celsius} \\
    Mon. -- Fri. 18:00--6:00 & 20 \si{\celsius}& 21 \si{\celsius}\\
    Sat. -- Sun. & 20 \si{\celsius}& 21 \si{\celsius}\\
    \end{tabular}
\end{table}

The formulation from Eq.~\eqref{eq_mpc} minimizes heating activity while also minimizing comfort band violations. To this end, the MPC controller computes an optimal sequence of future heating control actions over a prediction horizon $H_p$ using the model $f_\theta(\cdot)$. Once the optimal action trajectory is found, the first action is applied and passed to the building simulation. The resulting building response (observation) is returned to the MPC controller to repeat the optimization for the next trajectory. This process is repeated in a receding-horizon fashion. 

Model predictive control in buildings typically leads to improved comfort and reduced energy demand compared to standard control. However, the main challenge in MPC lies in obtaining an adequate model $f_\theta(\cdot)$ (Eq.~\eqref{eq_mpc_model}) of the target building's dynamics \cite{Drgona.2020}. In this work, we address this challenge using a TL model, which is pretrained on source building data and subsequently applied to control a target building. In the following, we first describe the model's pretraining and then its application to target building control.

\subsection{Generalized Transfer Learning Models}
\label{sec_pretraining}

\begin{figure}
	\centering
	\includegraphics[width=.99\columnwidth]{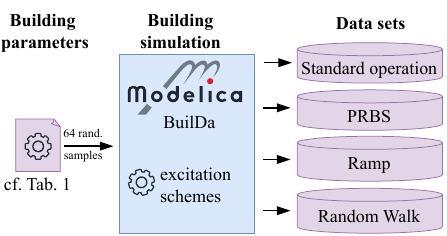}
	\caption{Data generation for pretraining the generalized models; displays blue part of Figure~\ref{fig_method}.}
	\label{fig_pretrain_method}
\end{figure}

For this study, we employ a generalized TL approach that uses multiple source buildings for pretraining, as it has been shown to outperform single-source-to-single-target TL models, models trained from scratch, and TSFMs \cite{koch2026thermalgems, raisch2025gentlgeneraltransferlearning, fm_nagy}. The generalized model is pretrained on data generated from the source building configurations described in Section~\ref{sec_data_gen} and Table~\ref{tab_permutations}. To simulate the sources, different control schemes can be employed to govern the buildings' thermal behavior. We refer to these as excitation schemes. We simulate the entire source dataset, consisting of the same 64 drawn buildings, once for each excitation scheme, as illustrated in Figure~\ref{fig_pretrain_method}. This allows us to identify which excitation scheme is best suited for pretraining the generalized model when the resulting model is later applied for MPC of an unseen target building. Each excitation scheme is introduced in the following.


\subsubsection{Excitation Schemes}
\label{sec_excite}

For the source data generation, we use the following excitation schemes:


\paragraph{Standard operation}
First, we employ standard operation, which is the basic approach for TL in buildings as in \cite{raisch2025gentlgeneraltransferlearning, pinto_sharing_2022, li_building_2024, koch2026thermalgems}. Standard operation reflects typical occupied-building conditions, where setpoints are given, and a standard rule-based or PI controller tracks them.
For the standard operating setup, we follow \cite{raisch2025gentlgeneraltransferlearning} and use the BuilDa-integrated proportional controller to maintain a randomly selected temperature setpoint. The setpoint is selected for 30\% of the simulations, with a constant setpoint between 20 \textdegree C and 24 \textdegree C, and a step size of 0.5 \textdegree C. The remaining 70\% apply a daytime setpoint within the same range combined with a night setback between 0.5 \textdegree C and 4 \textdegree C. This configuration captures both constant and day--night setpoint strategies under realistic operating conditions.

\paragraph{Pseudo Random Binary Sequence (PRBS)}
Second, we employ a pseudo random binary sequence (PRBS) signal, which is a widely used excitation method for building modeling \cite{bacher_identifying_2011, madsen1993short, 100c2e3b8bb24a83bab3ae769708cec8, knudsen_experimental_2021, SARTORI2023109149}. PRBS is a deterministic, binary signal that approximates band-limited white noise: it switches between on and off at pseudo-random times, with transitions restricted to discrete intervals $\lambda$. The integer $n$ sets the longest consecutive run in one state to $n\lambda$, and the sequence repeats with a maximum period of $N\lambda = (2^n - 1)\lambda$. We adopt the PRBS strategy from \cite{bacher_identifying_2011}, which combines short- and long-time constants to capture both air and envelope dynamics in buildings: $n=6$, $\lambda=20$~min for the short time constant, and $n=5$, $\lambda=210$~min for the long time constant.


\paragraph{Ramp excitation}

Similar to PRBS, the ramp excitation switches between on and off periods, but transitions linearly between the two extrema instead of switching instantaneously. Each transition between extrema ramps over a period uniformly sampled between 2 and 16 time steps. Once an extremum is reached, the signal remains constant for a holding period, sampled uniformly between 1 and 8 time steps, before the next transition begins. Unlike PRBS, this design ensures intermediate control values between 0 and 1 are represented in the training data, which can help a neural network learn the underlying state transitions more accurately.

\paragraph{Random walk}

\begin{figure}
	\centering
	\includegraphics[width=.9\columnwidth]{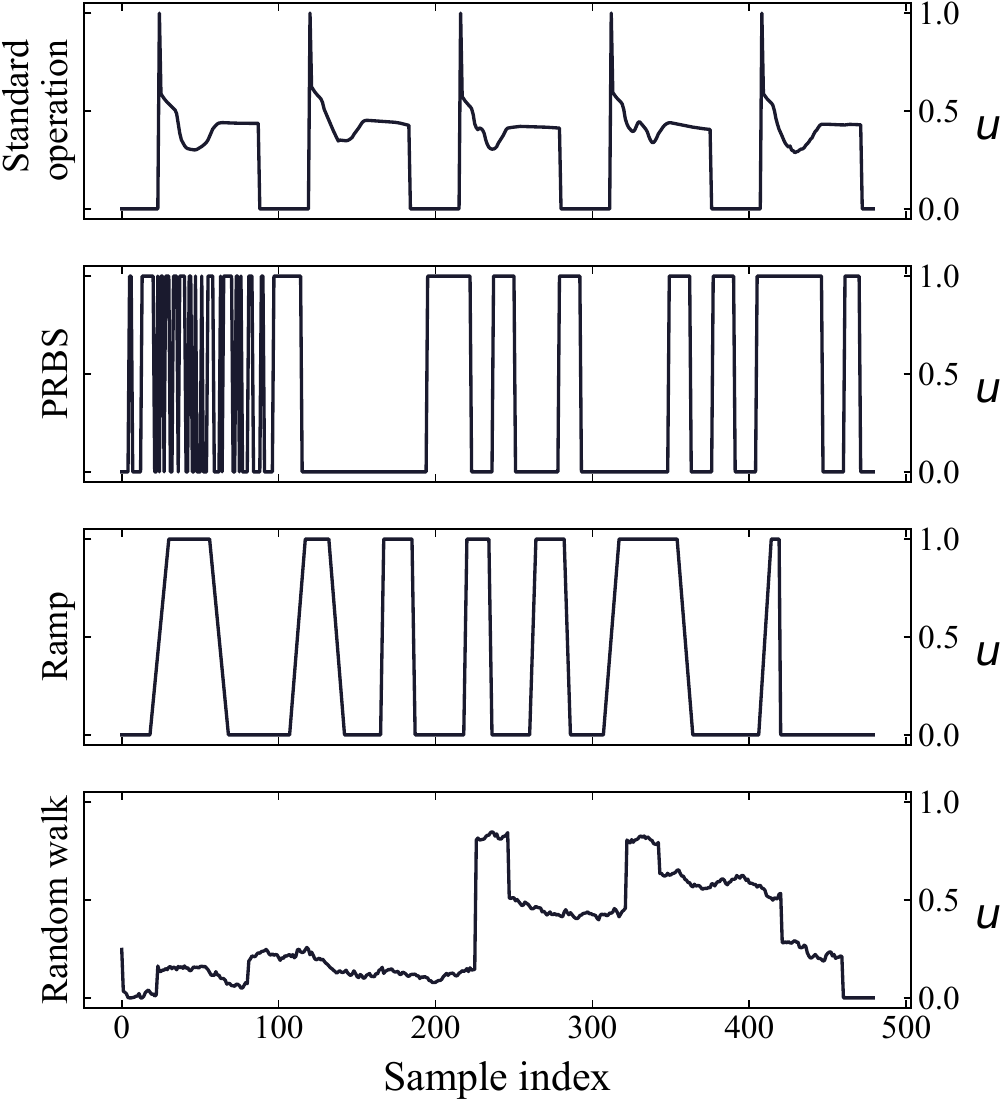}
	\caption{Different excitation schemes used for probing the building system. A control value of 1 corresponds to maximum heating power, and 0 corresponds to no power. An index step is 15 min.}
	\label{fig_excite_signals}
\end{figure}

Additionally, we consider a random walk excitation, which has been used in related system identification contexts to generate diverse, exploratory input signals for training data-driven models \cite{chen2026excite, EISCHENS2025110946, Haber2024RNN}. The control signal starts from a uniformly sampled initial value between $[0, 1]$ and is then updated at each time step by adding a Gaussian-distributed increment ($\mu$=0, $\sigma$=0.01), with the resulting value clipped to remain within $[0, 1]$. To avoid the signal stagnating within a fixed range, the walk is periodically reset to a new, uniformly sampled starting value after a randomly drawn interval of 4 to 96 time steps. Unlike PRBS and ramp excitation, the random walk follows no fixed switching pattern or predefined transition shape, providing a qualitatively different stochastic excitation. 
\\
\\
\noindent
An illustration of the excitation schemes can be found in Figure~\ref{fig_excite_signals}.
For each excitation strategy, the data are simulated over two winter periods, each spanning from 1 January to 30 March. This period was selected because it reflects the typical heating season in Central Europe, during which the heating signal $u$ is active. 

\subsubsection{Model Pretraining}
\label{sec_training}

Each source dataset from Section~\ref{sec_excite} is used to pretrain a generalized model. 
For control-oriented modeling, this model must capture the influence of the control action (input), i.e., the heating signal, on the indoor temperature (output). Measurable disturbances and auxiliary variables can additionally improve prediction accuracy. Following \cite{raisch2025gentlgeneraltransferlearning, RAISCH2026116868}, the neural network takes as input a lookback window of past measurements ($\mathbf{x}_{-lb:0}$), which are indoor temperature $T$, outdoor temperature $T_{out}$, direct and diffuse solar irradiation $Q_{dir}$ and $Q_{dif}$, and the heat source control signal $u_{in}$. Additionally, known future inputs are provided, comprising the weather forecast ($\mathbf{d}_{0:H_p-1}$) and the control trajectory $\mathbf{u}_{0:H_p-1}$, later generated by the MPC controller. A perfect weather forecast is assumed, as in \cite{STOFFEL2023100296, STOFFEL2023112709}. The output is the future indoor temperature $\hat{T}_{1:H_p}$ over the horizon $H_p$, so the governing Eq.~\eqref{eq_mpc_model} can be reformulated as follows:
\begin{equation}
\label{eq_TL_model}
    \hat{T}_{1:H_p} = f_\theta\!\left(\mathbf{x}_{-\text{lb}:0},\, \mathbf{d}_{0:H_p-1},\, \mathbf{u}_{0:H_p-1}\right).
\end{equation}
The enrolling index $k$ from Eq.~\eqref{eq_mpc_model} is omitted because Eq.~\eqref{eq_TL_model} always describes one full prediction horizon. We set the lookback $lb$ to 96 time steps, as in \cite{koch2026thermalgems}, and the horizon $H_p$ to 12 hours (48 time steps) to allow MPC optimization 12 hours ahead, as in \cite{STOFFEL2023100296, arroyo2022comparison}.

For modeling, we employ three neural network architectures: an MLP, a Long Short-Term Memory (LSTM) \cite{hochreiter1997lstm}, and a Transformer \cite{vaswani2017attention}. The MLP is widely used in building modeling and control due to its simplicity \cite{BUNNING2022118491, stoffel_real-life_2024, STOFFEL2023112709}. The LSTM is among the most widely used architectures for building dynamics and demonstrates strong performance in both short- and long-term prediction tasks \cite{raisch2025gentlgeneraltransferlearning, pinto_sharing_2022, li_building_2024}. The Transformer represents the current state of the art in time-series forecasting and has already demonstrated successful applications in modeling building thermal dynamics \cite{koch2026thermalgems, choi_performance_2023}. However, both LSTM and Transformer architectures are less frequently used in MPC-oriented control settings. The hyperparameters for each architecture are specified in Appendix~\ref{app_hypers}.

Each architecture is trained on each dataset from Section~\ref{sec_excite}, yielding 12 generalized models (see Figure~\ref{fig_method}). For training a generalized model, the corresponding source building time series is split into a training set (first winter) and a validation set (second winter).
The training data are segmented into training examples comprising the lookback and the horizon window, which are then shuffled across all 64 source buildings and used to train the neural network in random order. The same segmentation process is applied to the validation set, which is used to select the best-performing model checkpoint across training epochs, thereby avoiding overfitting to the training set. No separate test set is required for the source dataset, as evaluation is performed on the target buildings.

All models were trained on an AIME T600 workstation equipped with an NVIDIA RTX A6000 GPU (48\,GB), an AMD Threadripper Pro 5995WX CPU, and 512\,GB of RAM. On average, training took 17\,min for the MLP, 5\,h\,25\,min for the LSTM, and 51\,min for the Transformer. The substantially longer training time for the LSTM is due to its sequential unrolling over time, which limits parallelization compared to the MLP and Transformer architectures.


\subsubsection{Target Application of Generalized Models}
\label{sec_zero}

Once pretrained, the generalized models can be applied to an unseen target building. In TL settings, this can be done either in a zero-shot manner or through fine-tuning using target-specific data. Zero-shot refers to applying the pretrained model directly to a target building without any fine-tuning, using only the lookback window (context) of past measurements as input. Based on this lookback window, along with a weather forecast and the future control trajectory, the model predicts the indoor temperature of the target. This prediction model can then be used for MPC in the target building. Fine-tuning, on the other hand, first uses target-specific measurement data to adjust the pretrained model's weights before employing it for target predictions. However, \cite{koch2026thermalgems} reports that fine-tuning does not necessarily improve prediction accuracy and therefore recommends zero-shot inference as a simple and effective approach. Beyond this, zero-shot application avoids any additional data collection in the target, thereby reducing operational cost. It also establishes a conservative lower bound on TL performance, since fine-tuning should, in principle, only improve upon it. Accordingly, we evaluate target control and prediction performance under the zero-shot setting.

Having obtained a pretrained model for zero-shot prediction, we can now incorporate it into the MPC optimization by replacing Eq.~\eqref{eq_mpc_model} with Eq.~\eqref{eq_TL_model}. Based on this model, the MPC computes a control sequence over the prediction horizon by minimizing the cost function. We set the cost function parameters from Eq.~\eqref{eq_mpc_obj} to $w_0 = 1$, $w_1 = 1$, and $w_2 = 10$. Further discussion on weight selection, along with additional weighting configurations, is presented in Appendix~\ref{app_weighting}. 
For the neural-network models, the pretrained network defines a non-convex but differentiable mapping from the control sequence to the predicted temperature trajectory. This allows the control sequence to be optimized directly via gradient descent: the network weights remain frozen, while gradients of the cost with respect to the control sequence are obtained through automatic differentiation in PyTorch \cite{paszke2019pytorch}. The optimization is run for 200 iterations, and the best-performing control sequence found with regard to the objective function is used for target control. This approach is similar to \cite{pmlrjain20a}, who use the same gradient-based strategy but implemented in TensorFlow.

\subsection{Evaluation Setup and Baselines}
\label{sec_eval}


For evaluation, we use the distinct set of 32 target buildings described in Section~\ref{sec_data_gen}, assessing both MPC and prediction performance. For control, each target is controlled via MPC as described in Section~\ref{sec_target_control}. 
MPC control of the targets starts in January, as this accounts for the heating period. Before, we assume standard operation of the target buildings, as explained in Section~\ref{sec_excite}.
We evaluate control performance over a one-month period, as assessing all model and excitation-strategy combinations over a longer period is computationally expensive. For a selected subset of experiments, we also report results over a ten-week period. 

We additionally assess the models' predictive accuracy on the target buildings to account for the accuracy paradox (prediction--control performance gap), as introduced in Section~\ref{sec_intro}. For this evaluation, we consider two test sets. The first, referred to as the \textbf{standard operation test set (1)}, reflects target building operation under standard operation as introduced in Section~\ref{sec_excite} and enables comparison with previous TL studies \cite{raisch2025gentlgeneraltransferlearning, pinto_sharing_2022, li_building_2024}. The second, referred to as the \textbf{MPC-based test set (2)}, consists of data obtained during actual MPC control and allows us to assess prediction performance under real MPC operation. Both testing protocols are evaluated over the same evaluation window used for the control performance assessment above.
\\
\\
\noindent
For benchmarking the generalized models, we consider three baselines, which are introduced below:

\paragraph{Unexcited pretraining}
As a first baseline, we compare our excitation-based pretraining strategies against pretraining on standard operational source data, as introduced in Section~\ref{sec_excite}. This baseline reflects the predominant approach used in existing TL literature \cite{raisch2025gentlgeneraltransferlearning, pinto_sharing_2022, li_building_2024} and allows us to directly assess the benefit of dedicated excitation strategies for downstream MPC performance.

\paragraph{Linear model}
Additionally, we employ a linear model, as it has been identified as the strongest candidate for MPC applications in buildings \cite{stoffel_real-life_2024, knudsen_experimental_2021, BUNNING2022118491}. Linear models are particularly advantageous in MPC settings, as they yield computationally efficient, convex optimization problems. Additionally, their linear structure leads to beneficial extrapolation behavior. We adopt the linear model from \cite{stoffel_real-life_2024}, as it outperformed all other ML-based models for MPC in their study. The model is set up as an online multi-linear regression model and uses the same feature set as the neural-network models from Section~\ref{sec_pretraining}. Following \cite{stoffel_real-life_2024}, the model includes 6 lags of the heat source control signal $u_{in}$, 3 lags of the outdoor temperature $T_{out}$, 4 lags of each direct and diffuse solar irradiation $Q_{dir}$ and $Q_{dif}$, and 1 lag for the indoor temperature $T$. Notably, the authors report that the linear model showed little sensitivity to feature and lag selection. Thereby, Eq.~\eqref{eq_mpc_model} is rewritten to:
\begin{equation}
    T_{k+1} = \beta_0 + \sum_i \beta_i z_i.
\end{equation}
Here, $z_i$ denotes the input variables and their lags, and $\beta_i$ the regression coefficients. When enrolling $T_{k+1}$ over the horizon $H_p$, the same future weather forecast as for the generalized models is used (see Section~\ref{sec_training}).

Following the same setup of \cite{stoffel_real-life_2024}, the linear model is trained online on 12 days of past data, retrained daily throughout the MPC evaluation. Longer training periods do not necessarily improve linear models' performance \cite{raisch2026transferlearningneuralparameter, BLUM2019410}. As the MPC evaluation takes place in January, the initial training samples are drawn from the last 12 days of December (standard operation). The online setup renders our comparison between the generalized models and the linear model conceptually similar to that of \cite{TSFM_zero}, who compared time series foundation models against linear baselines trained on the lookback window (2--7 days). However, we grant the linear baseline 12 days of past data compared to the 1-day lookback used for the generalized models. 

The optimization problem arising from the linear formulation is a convex quadratic program because the model is linear in the control inputs and the cost function is quadratic. This allows it to be solved in closed form, rather than requiring the iterative gradient-based optimization used for the neural network models. 

\paragraph{PI controller}
As an additional baseline, we use the proportional-integral (PI) controller provided by BuilDa. PI controllers represent a standard, model-free control approach widely used in real buildings. The PI controller regulates the heating signal $u$ based on the error between a temperature setpoint $T^{\text{set,PI}}$ and the measured indoor temperature.
For the reference setpoint, we employ the lower bound of the temperature band from Table~\ref{tab_Tset} ($T^{\text{set,PI}}=T^{\text{low}}$). This encourages minimal heating activity while satisfying comfort requirements.
Unlike the MPC approaches, the PI controller does not rely on a predictive model of the building dynamics and, therefore, can only operate reactively.

\section{Experiments}
\label{sec_results}

In this section, we present the experiments and results of our study. For all experiments, we assess control performance or prediction accuracy across the total set of 32 target buildings. 

In Section~\ref{sec_control_ex}, we evaluate the ability of the generalized models for target MPC. Accordingly, we evaluate the generalized model pretrained on standard operation data against the models pretrained on excitation-based data across all architectures.
Following, we assess the prediction--control performance gap. For this evaluation, we compare the control performance of individual target buildings against the generalized models' prediction accuracy on both testing protocols, the standard operation and MPC-based test sets.

Next, in Section~\ref{sec_bench_ex}, we compare the best-performing generalized model from Section~\ref{sec_control_ex} to the benchmark control approaches, described in Section~\ref{sec_eval}. 


\subsection{Generalized Models}
\label{sec_control_ex}

\begin{figure*}
	\centering
	\includegraphics[width=.9\textwidth]{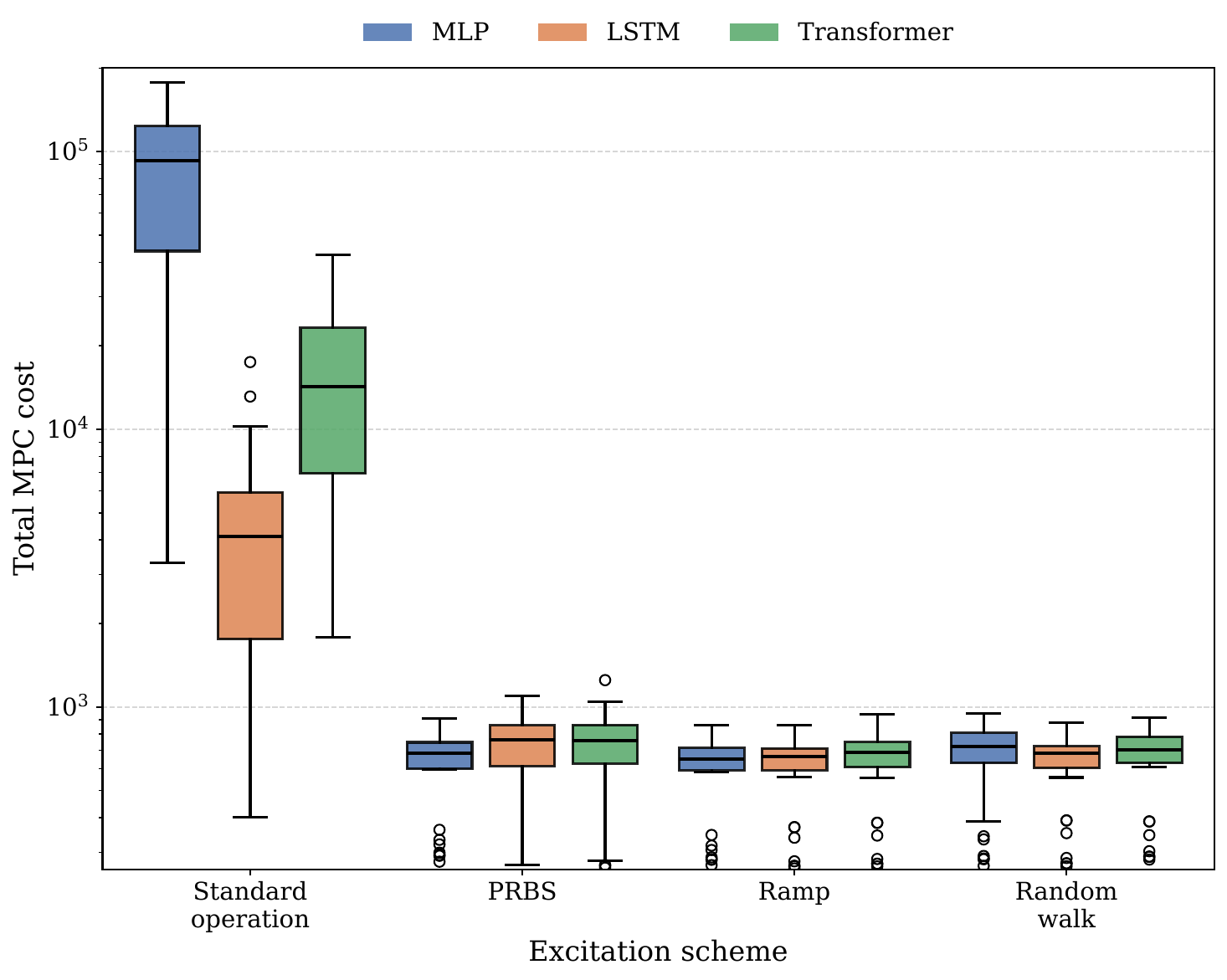}
	\caption{Total MPC cost in logarithmic scale across the 32 target buildings for generalized models with different architectures and excitation strategies. Results are shown as box plots, indicating the median, interquartile range, and outliers. The x-axis denotes the excitation strategy used for pretraining.}
	\label{ex_mpc_compare}
\end{figure*}

We first compare all 12 generalized models, considering different architectures and excitation schemes, on their target control performance. The resulting total MPC costs across all target buildings are shown in Figure~\ref{ex_mpc_compare} as box plots (Table~\ref{tab_cost_compare} reports the corresponding mean values).
The plot shows that models trained on standard operational data lead to substantial costs during MPC control of the targets. The MLP incurs the highest cost, with a median close to $10^5$ and substantial spread, followed by the Transformer, while the LSTM outperforms both. On the other hand, models trained on excitation-based data lead to substantially better control performance. For excitation-based pretraining, architectural differences vanish: MLP, LSTM, and Transformer achieve comparable median costs and similarly narrow interquartile ranges. Among the excitation strategies, ramp excitation appears slightly superior, achieving the lowest median, variance, and mean cost (see Table~\ref{tab_cost_compare}), followed by the random walk, with both maintaining costs consistently below $10^{3}$. PRBS shows a larger variance and slightly higher median costs than the other two excitation schemes.
\\
\begin{figure*}
	\centering
	\includegraphics[width=.99\textwidth]{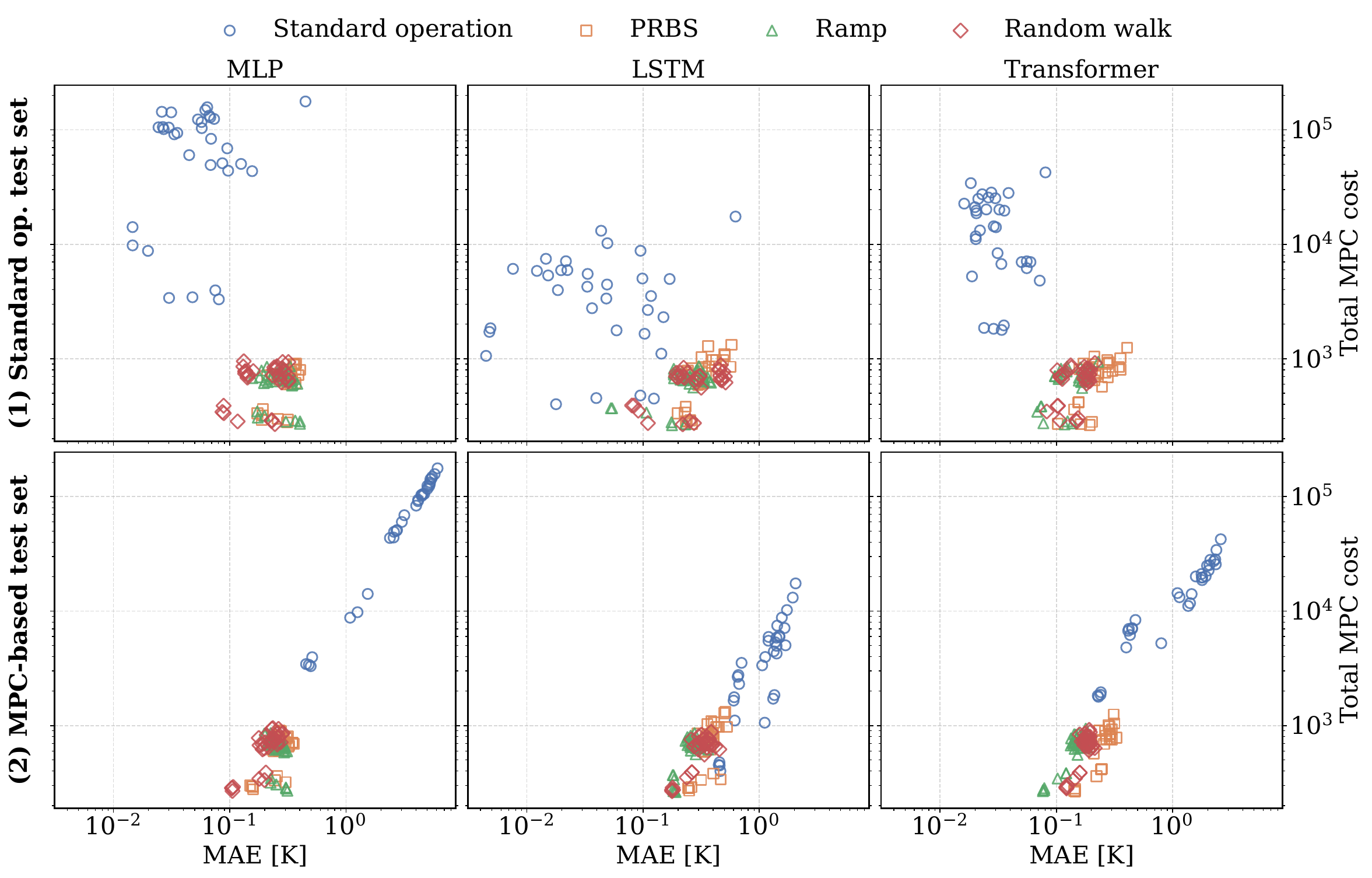}
	\caption{Relationship between prediction accuracy (1h ahead, MAE) and total MPC cost for each architecture, colored by excitation strategies (both axes on a logarithmic scale). For prediction accuracy, the standard operation test set (1) and the MPC-based test set (2) are used.}
	\label{ex_scatter_cost_pred}
\end{figure*}

\noindent
Next, we assess the relationship between each model's prediction accuracy for a specific target building and its resulting MPC performance in Figure~\ref{ex_scatter_cost_pred}.
Two prediction accuracy protocols are shown: the standard operation test set (1) and the MPC-based test set (2). For both protocols, the mean absolute error (MAE) is reported. Each marker corresponds to one target building, with the marker shape and color indicating the excitation scheme used for pretraining.

On the standard operation test set (1), generalized models pretrained on standard operational source data (blue circles) achieve the lowest prediction error across all architectures with MAE values clustered around $0.01-0.1$~K. However, when the same models are evaluated in terms of total MPC cost, they exhibit the highest costs compared to all other models, highlighting the accuracy paradox. Similarly, when evaluated for prediction accuracy on the MPC-based test set (2), these models exhibit the highest prediction errors. In contrast, models pretrained on any of the excitation strategies (PRBS, ramp, random walk) retain the same level of prediction accuracy when evaluated on the standard operation test set (1) or the MPC-based test set (2). Excitation-based models cluster tightly at low MPC cost and low prediction accuracy around $0.1-0.6$~K, with no strong separation among the individual excitation strategies. 
Generally, a roughly linear correlation between prediction accuracy and control performance can be observed for the MPC-based test set (2). This is particularly visible for the generalized models pretrained on standard operational source data. 
With regard to both control and prediction performance, this plot underlines the strong results of ramp excitation, clustered bottom-left, closely followed by random walk, with PRBS performing slightly worse. Also, the plot shows that good test-set performance on the standard operation test set (1) alone does not necessarily indicate good control performance.

\begin{table*}[thb]
\centering
\caption{MPC costs, prediction accuracy, and MPC optimization time averaged across target buildings, by model architecture and excitation scheme. Prediction accuracy (MAE) in K is assessed on the respective test set. Bold values indicate the lowest value per row.}
\label{tab_cost_compare}
\resizebox{\textwidth}{!}{%
\begin{tabular}{c l cccc cccc cccc}
\toprule
& & \multicolumn{4}{c}{MLP} & \multicolumn{4}{c}{LSTM} & \multicolumn{4}{c}{Transformer} \\
\cmidrule(lr){3-6} \cmidrule(lr){7-10} \cmidrule(lr){11-14}
& & Standard & PRBS & Ramp & Rand. walk & Standard & PRBS & Ramp & Rand. walk & Standard & PRBS & Ramp & Rand. walk \\
\midrule
\multirow{3}{*}{\rotatebox{90}{Cost}}
& Total cost   & 81198.20 & 636.13 & \textbf{607.18} & 669.65 & 4596.60 & 737.15 & 610.36 & 624.83 & 15703.28 & 716.33 & 636.52 & 650.29 \\
& Discomfort & 80915.29 & 45.54  & \textbf{30.22}  & 109.90 & 4076.26 & 142.78 & 45.78  & 65.01  & 15153.18 & 147.99 & 76.13  & 89.00  \\
& Heating activity  & \textbf{282.91} & 590.59 & 576.96 & 559.75 & 520.34 & 594.37 & 564.58 & 559.82 & 550.10 & 568.34 & 560.39 & 561.29 \\
\midrule
\multirow{2}{*}{\rotatebox{90}{MAE}}
& Standard op. test set & 0.069 & 0.302 & 0.273 & 0.213 & 0.075 & 0.337 & 0.242 & 0.299 & \textbf{0.033} & 0.227 & 0.140 & 0.152 \\
& MPC-based test set & 3.564 & 0.265 & 0.257 & 0.214 & 1.144 & 0.364 & 0.252 & 0.314 & 1.313 & 0.240 & \textbf{0.140} & 0.175 \\
\midrule
& Optimization time & \multicolumn{4}{c}{\textbf{0.822 s}} & \multicolumn{4}{c}{9.476 s} & \multicolumn{4}{c}{1.539 s} \\
\bottomrule
\end{tabular}%
}
\end{table*}

For a better comparison, Table~\ref{tab_cost_compare} provides mean costs and mean prediction accuracies over all targets for each model. Here, the MLP pretrained on ramp-excitation data achieves the best comfort and total cost. Based on the relative performance improvement\footnote{(cost$_{\text{benchmark}}$ $-$ cost$_{\text{model}}$)$/$cost$_{\text{benchmark}}$.}, the MLP based on ramp excitation achieves an average 86.79\% improvement in control performance over the best generalized model trained on standard operational source data (LSTM).
As shown in Figure~\ref{ex_scatter_cost_pred}, the MAE of the prediction accuracy on the MPC-based test set (2) correlates linearly with MPC performance. Hence, the difference between the MAE on the standard operation test set (1) and the MAE on the MPC-based test set (2) can serve as an indicator of the prediction--control performance gap. For the generalized model pretrained on standard operational data, we observe this prediction gap between (1) and (2) to range between 1.07~K and 3.50~K, compared to a range of 0.0~K to 0.037~K for the excitation-based generalized models. The gap is therefore substantially smaller for the excitation-based models. 

In terms of computational cost, the update time is independent of the excitation strategy and depends only on the architecture: the MLP achieves the fastest average time per optimization step at 0.82 s, followed by the Transformer at 1.54 s, while the LSTM is considerably slower at 9.48 s per update, due to its sequential, recurrent structure.

Following this, we continue with the MLP model pretrained on ramp-excitation data, as this configuration achieves both the lowest MPC cost and the shortest MPC optimization time.

\subsection{Comparison to Benchmarks}
\label{sec_bench_ex}

In this section, we compare the best-performing generalized model (MLP trained on ramp excitation) to the standard baseline controllers from Section~\ref{sec_eval}. For the unexcited baseline, we use the LSTM, as it performed best within its category (generalized models pretrained on standard operational data) in the previous experiments.

\begin{figure}[]
	\centering
	\includegraphics[width=0.99\columnwidth]{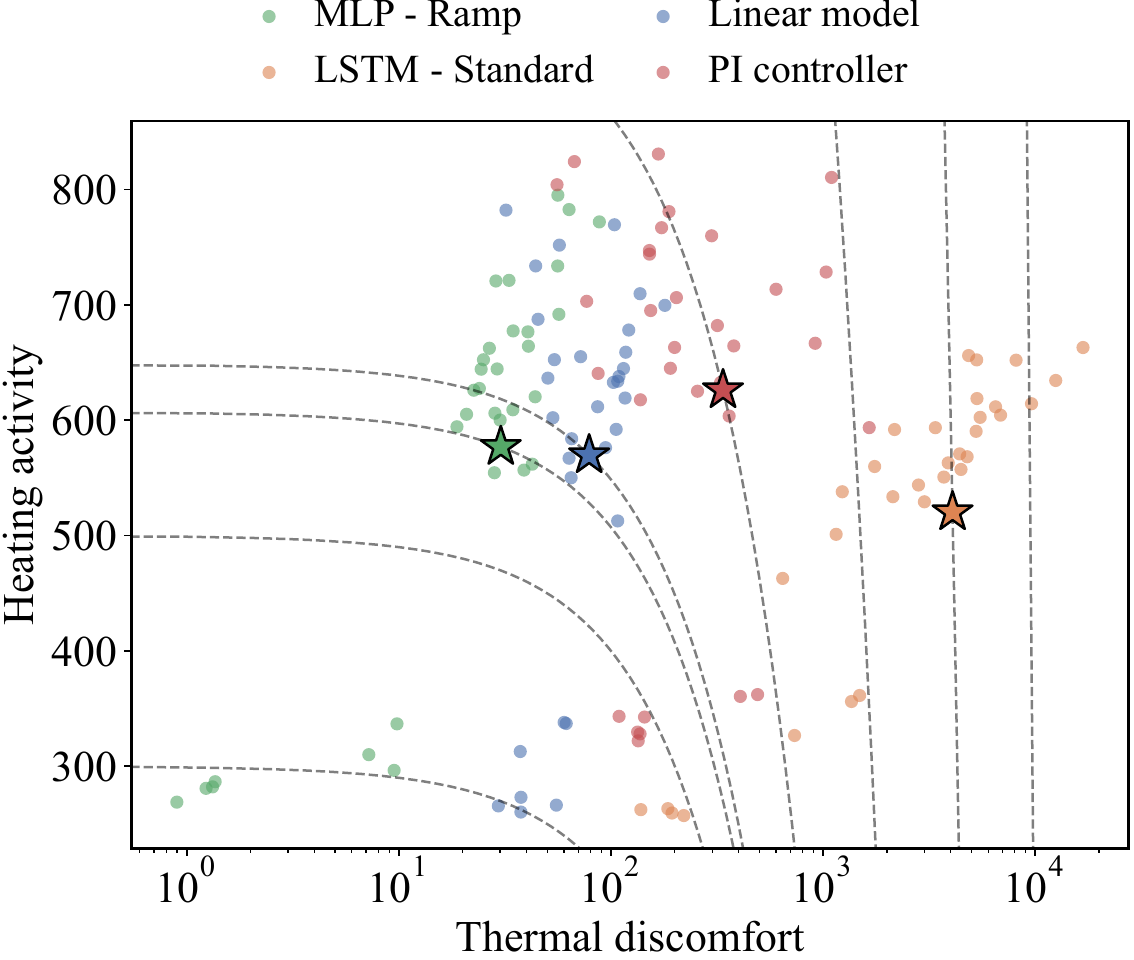}
	\caption{Scatter plot of thermal discomfort (logarithmic scale) versus heating activity across target buildings. Dashed lines indicate constant total cost (the sum of thermal discomfort and heating activity); star markers denote the mean cost of each method, with a dashed line passing through each mean.}
	\label{ex_benchmark}
\end{figure}

Figure~\ref{ex_benchmark} shows the control performance of each method as a scatter plot across the 32 target buildings. The two axes correspond to the two MPC objectives: thermal discomfort (log scale) and heating activity (linear scale). The star marks the average cost for each method, and the dotted lines indicate constant total cost. Overall, the excitation-based MLP achieves the lowest average total cost, with buildings clustered at low thermal discomfort (below $10^2$) and moderate heating activity.
The linear model ranks second, with slightly lower average heating activity than the MLP, but larger comfort violations (see Table~\ref{tab_setpoint_cost}). The PI controller ranks third, with higher and more variable heating activity and comfort cost than the other two methods. The LSTM, pretrained on standard operational data, incurs the highest control costs overall.

For a better comparison, Table~\ref{tab_setpoint_cost} shows the average cost per method.
On average, the MLP based on ramp excitation outperforms the linear model by 6.4\% and the PI controller by 36.9\% using the relative performance improvement introduced above.
For a building-level evaluation, we refer to Appendix~\ref{app_build_level}, which shows that the generalized model pretrained on ramp excitation outperforms the linear model on 25 of the 32 target buildings, and outperforms both the PI controller and the generalized model pretrained on standard operational data across all 32 buildings.
\\

\begin{figure}
	\centering
	\includegraphics[width=.99\columnwidth]{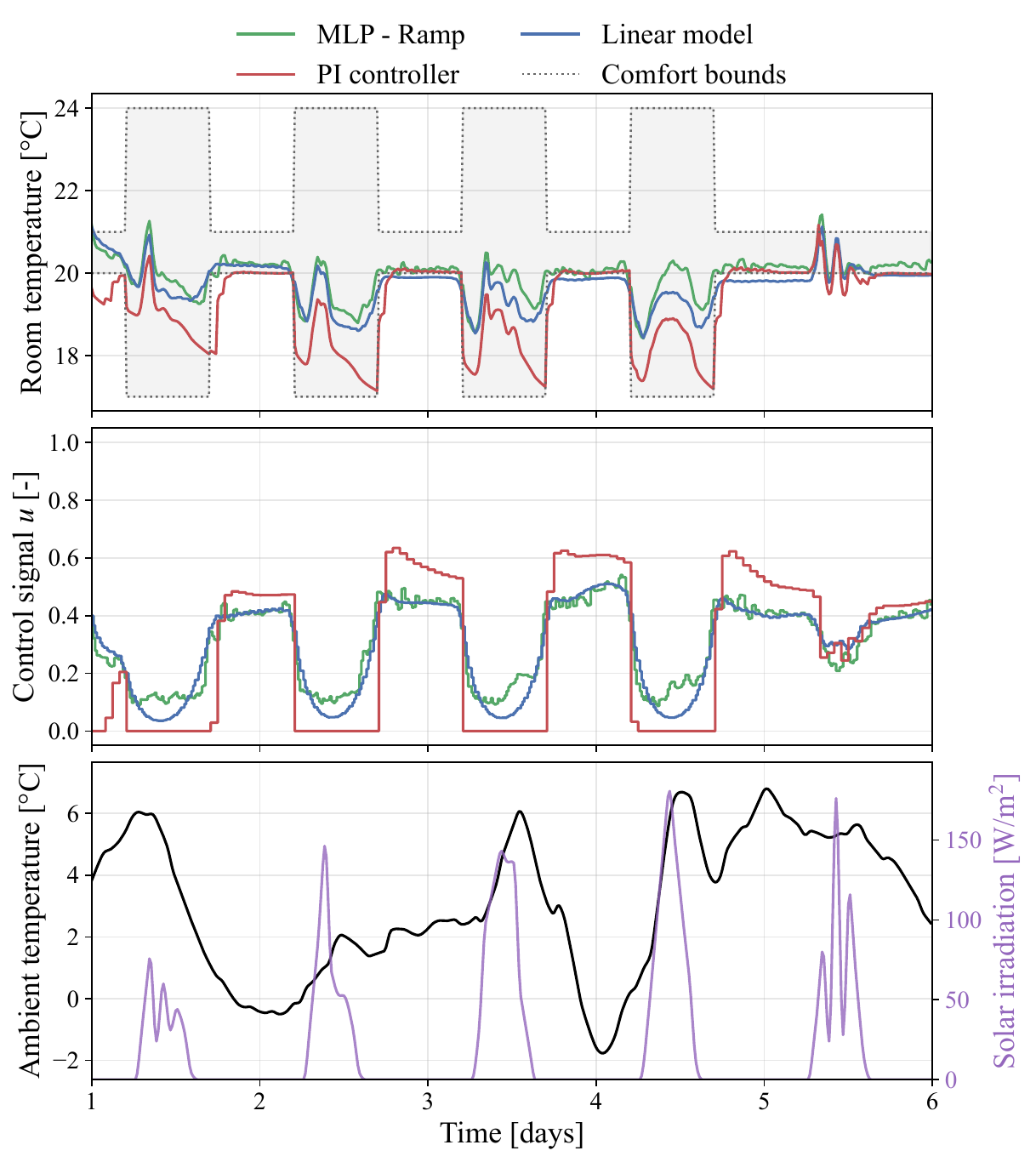}
	\caption{Control trajectories for the first days of January for one example building. Also given are the outside temperature and solar irradiation. The shaded area in the first subplot displays predictions of the MLP and linear model.}
	\label{ex_mpc_ramp_mlp}
\end{figure}

\begin{table}
\centering
\caption{Average total cost, discomfort, and heating activity across all 32 targets per approach, with best values indicated in bold.}
\label{tab_setpoint_cost}
\resizebox{\columnwidth}{!}{%
\begin{tabular}{l|rrr}
\toprule
 & Total cost & Discomfort & Heating activity \\
 \hline
 MLP - Ramp & \textbf{607.2} & \textbf{30.2} & 577.0 \\
LSTM - Standard & 4596.60 & 4076.26 & \textbf{520.34} \\
Linear model & 648.6 & 78.9 & 569.7 \\
PI controller & 963.5 & 337.3 & 626.2 \\
\bottomrule
\end{tabular}
}
\end{table}
\noindent
Next, Figure~\ref{ex_mpc_ramp_mlp} shows the temperature and control trajectories during MPC operation for a target building. This plot compares MPC based on the generalized ramp-excitation model, linear-model-based MPC, and the PI controller. The LSTM with standard operation excitation is omitted for visual clarity.
The generalized MLP-based MPC reliably keeps the room temperature within the comfort bounds while modulating the control signal in response to ambient temperature and solar irradiation. The linear model shows a similar overall trend but occasionally deviates from the comfort bounds. 
The PI controller, in contrast, tracks its fixed lower-bound setpoint closely but does not anticipate upcoming weather conditions or setpoint changes, resulting in a control signal that reacts only after temperature changes have already occurred rather than in advance.
\\

\noindent
In the Appendix, we provide additional results. Appendix~\ref{sec_pred_lin} presents a prediction-accuracy assessment for the linear model for both the standard operation test set (1) and the MPC-based test set (2), where we observe that linear models generalize substantially better than the LSTM pretrained on standard operation data, but worse compared to the MLP pretrained on ramp excitation. Appendix~\ref{app_3M_control} additionally reports results over a ten-week control period, showing consistently better performance of the MLP on a weekly basis.

\section{Discussion}
\label{sec_discuss}
This paper investigates pretrained generalized TL models for MPC of unseen target buildings in zero-shot mode. To this end, we compare the classic TL approach of pretraining a generalized model on standard operational data, i.e., comfort-setpoint-driven operations, against the benefit of using excitation-based data for pretraining. For the excitation strategies, we consider a pseudo random binary sequence (PRBS), a ramp signal, and a random walk. Additionally, three neural network architectures are assessed for pretraining: a multi-layer perceptron (MLP), a long short-term memory (LSTM) network, and a Transformer. All generalized models are evaluated for MPC performance and prediction accuracy in 32 different target buildings.
Finally, we compare the best-performing generalized model to standard baseline controllers, namely an online linear model-based MPC and a PI controller. 

\subsection{Principal Findings}

In the following, we present the main findings of this work and conclude by answering the research questions posed in the Introduction.

\paragraph{TL models for control}
Generalized models trained on standard operational data achieve good prediction accuracy on the standard operation test set, consistent with prior studies \cite{raisch2025gentlgeneraltransferlearning, koch2026thermalgems}. However, when evaluated for control, these models yield poor MPC performance across all target buildings. Pretraining on excitation-based data, by contrast, substantially improves control performance, reducing control cost by around 86\% while maintaining an acceptable prediction error generally below 0.3~K. This leads to a key takeaway: excitation strategies during the pretraining of generalized models are highly relevant to downstream control performance, whereas pretraining on standard operational data is insufficient for satisfactory control. 
Furthermore, these results highlight that prediction accuracy on a standard test set alone does not indicate good control performance, as found before \cite{martin_poster, BLUM2019410, stoffel_real-life_2024}.
We therefore cannot recommend the prediction accuracy assessment used in previous TL studies for estimating control effectiveness. This finding, though, is based on the generalized model approach and consequences may therefore differ for single-source-to-single-target TL. Nonetheless, prior studies in this domain, such as \cite{li_building_2024, pinto2022transfer, DOU2025113341, cho2024application, chen2020transfer, chaudhary2025transfer, fm_nagy}, should be reconsidered for target control rather than prediction alone.

\paragraph{Prediction and MPC performance gap} 
As an indicator of the prediction--control performance gap, we use the difference in prediction accuracy between the standard operation test set (1) and the MPC-based test set (2), since accuracy on the MPC-based test set (2) correlates approximately linearly with total MPC cost. Models pretrained on standard operational data achieve good accuracy on the standard operation test set (1), but this degrades substantially on the MPC-based test set (2), with a gap of 1.07--3.50~K, while also achieving poor control performance. Models pretrained on excitation-based data, by contrast, retain comparable accuracy on both test sets, with a gap of only 0.0--0.037~K -- indicating that excitation-based pretraining narrows the prediction--control performance gap identified as an open challenge in the literature \cite{martin_poster}.

\paragraph{Comparison to benchmarks}

For comparison against the benchmark control approaches, we use the MLP pretrained on ramp-excitation data, which achieves the best overall control performance among all evaluated models. Across all target buildings, we observe improvements over the linear model of 6.4\% and over the PI controller of 36.9\%, respectively. 
The PI controller generally achieves acceptable results, as it tracks a fixed reference temperature closely, but is limited to reactively following setpoints, rather than providing predictive control.
For the linear model benchmark, we observe good predictive performance overall. However, the prediction gap between the two sets is 0.56~K, which is higher than that of the MLP trained on ramp excitation (0.016~K). Further, the linear model achieves good MPC results and remains a strong benchmark, consistent with the findings of \cite{stoffel_real-life_2024, BUNNING2022118491, knudsen_experimental_2021}. Stoffel et al. \cite{stoffel_real-life_2024} report comfort and heating activity savings of 86.0\% and 5.5\% for their linear model over the PI controller, which is comparable to our linear model's savings of 76.6\% and 9.0\% over our PI controller. However, unlike generalized models, which are pretrained offline and applied directly to unseen target buildings in a zero-shot manner, the linear model relies on online learning and therefore must be retrained daily. This requires continuous collection and storage of building-specific operational data, along with a recurring retraining procedure, adding practical overhead that the zero-shot TL approach avoids entirely.

\paragraph{Excitation techniques}

Excitation-based pretraining substantially improves control performance in target buildings compared to unexcited pretraining. Among the excitation schemes themselves, we observe only small differences in downstream control performance. Ramp excitation leads to the overall best performance, followed by the random walk, while PRBS achieves the weakest, though still satisfactory, results for MPC. While PRBS is commonly used for RC-parameter estimation in gray-box building models \cite{bacher_identifying_2011, knudsen_experimental_2021}, this excitation approach primarily activates certain frequencies in the output temperature data. However, PRBS only switches the input control signal between 0 and 1, whereas neural networks instead require diverse input and output data across the relevant state-action space \cite{chen2026excite, EISCHENS2025110946, Haber2024RNN}. This likely explains why the ramp signal (and also the random walk), which explicitly covers intermediate control values, achieves better performance than PRBS.

\paragraph{Neural architectures}

Regarding the architecture comparison, the LSTM achieves the best control performance among the three architectures when pretrained on standard operational data. However, because standard operational pretraining does not yield acceptable overall control performance, this finding is of limited practical relevance. For excitation-based pretraining, the differences among the architectures are small, making all three neural networks suitable for the proposed method. Nonetheless, the MLP performed slightly better than the LSTM and the Transformer. A further advantage of the MLP is its short pretraining time of 17 minutes and MPC optimization time of 0.8 seconds per update step, corresponding to roughly a 2-times speed-up over the Transformer and an 11-times speed-up over the LSTM. 

\paragraph{Implications for the research questions}

In summary, these key findings help us answer the research questions stated in the Introduction: generalized models pretrained on standard operational data are largely unsuitable for downstream MPC applications, as they cannot compete with any of the benchmarking methods. Their high prediction accuracy does not translate into good control performance, confirming the accuracy paradox. In contrast, incorporating excitation schemes during pretraining substantially improves control performance in the target building, effectively narrowing the prediction--control performance gap. Finally, generalized models pretrained on excitation data can outperform linear models in downstream control applications without any target-specific fine-tuning. 

With these findings, we take a step toward addressing the barriers to data-driven MPC in buildings, as explained in the introduction: the need to repeat data collection and model training for individual buildings. The excitation-based generalized model helps mitigate these barriers, as it can be applied directly to a target building without any target-specific data, model training, or fine-tuning.

\subsection{Limitations and Future Work}
\label{sec_limits}

One limitation of this work is the exclusive evaluation on simulated target buildings. While the source data generation is intended to be at the simulation level to enable excitation-based data generation across many buildings, target control would ideally be assessed on real buildings. This introduces a simulation-to-reality challenge when applying a synthetically pretrained generalized model to real target buildings, as, for example, investigated in \cite{DOU2025113341} for temperature prediction in buildings. As a first step, our study focuses on evaluating downstream control performance across multiple simulated target buildings. Future research is planned to further validate the pretrained models on a real-world test-site building.

Additionally, this study is limited to single-zone residential buildings in Central Europe and considers only heating-dominated operation. Accordingly, the generalized model proposed in this paper is suited mainly for this domain. Nevertheless, the presented approach is not conceptually limited to this setting. It could be extended toward a single foundation model approach for MPC across all buildings, generalizing across different building types, multi-zone buildings, and locations worldwide. An alternative is to generate a collection of domain-specific models, as done in this paper, each trained for a different building domain. Extending the presented methodology to multi-zone and non-residential buildings, to combined heating and cooling operation, therefore, remains an important and promising direction for future work.

A further aspect, which we consider more of a design choice than a limitation, is our evaluation in a zero-shot manner. A well-executed fine-tuning approach could further improve performance beyond zero-shot, suggesting our reported results may in fact be conservative. However, this remains an open challenge: \cite{koch2026thermalgems} reports that naive fine-tuning, i.e., fine-tuning based solely on weight initialization, does not necessarily improve accuracy over zero-shot prediction, while \cite{pinto_sharing_2022} recommends weight initialization over freezing selected layers, and \cite{RAISCH2026116868} shows that continual fine-tuning over time appears important. Future work should therefore systematically evaluate fine-tuning strategies with respect to their closed-loop control performance, without unlearning the benefits acquired through excitation-based pretraining.

Another future research direction is the hyperparameter selection of the generalized models. 
We deliberately avoid typical hyperparameter tuning, as this would optimize performance on source validation data, which is not indicative of downstream MPC performance in a target. This is in line with earlier studies, which have similarly used self-defined MLP models rather than tuning hyperparameters \cite{stoffel_real-life_2024, STOFFEL2023112709}. Future research could therefore optimize hyperparameters directly with respect to control performance, potentially based on \cite{amos2021optnetdifferentiableoptimizationlayer}.

\section{Conclusion}
\label{sec_conclusion}

This paper investigates generalized transfer learning (TL) models for model predictive control (MPC) of unseen target buildings. These models are pretrained on multiple source buildings and applied to a target building via zero-shot.
We show that pretraining on standard operational data results in high control costs across target buildings. In contrast, pretraining on excitation-based source data reduces these costs by an average of 86.8\%. Moreover, excitation-based pretraining outperforms an online linear model-based MPC that uses target-specific data by an average of 6.4\%, and a PI controller by 36.9\%. By transferring robustly to a wide range of buildings without target-specific data collection or modeling effort, our approach contributes to reducing MPC setup cost, supporting the wider adoption of energy-efficient control in buildings.

\printcredits

\section*{Declaration of competing interests}

The authors declare that they have no known competing financial interests or personal relationships that could have appeared to influence the work reported in this paper.

\section*{Declaration of generative AI and AI-assisted technologies in the manuscript preparation process}

During the preparation of this work, the authors used Claude (Anthropic) in order to improve language, clarity, and sentence structure of the manuscript text. After using this tool, the authors reviewed and edited the content as needed. The authors take full responsibility for the content of the published article.

\section*{Data availability}

The code and data used to perform the experiments in this study are available at \cite{github_exciting}.

\FloatBarrier
\appendix
\counterwithin{table}{section}
\counterwithin{figure}{section}
\section{Hyperparameter Selection}
\label{app_hypers}

For the different architectures, we adopt the hyperparameters from \cite{koch2026thermalgems}, as they performed tuning for a similar dataset and task. Table~\ref{tab_gm_hyperparams} summarizes the hyperparameters. The Transformer consists of three transformer encoder blocks followed by two dense layers serving as the decoder. To cope with the inherent temporal invariance of the transformer architecture, we added a cosine positional embedding to the input features as proposed in \cite{vaswani2017attention}. Since \cite{koch2026thermalgems} does not include an MLP architecture, we instead select well-performing hyperparameters for the MLP based on preliminary experiments. 

For our study, standard hyperparameter tuning is of secondary importance, as such tuning would optimize performance solely based on the validation-set performance of the source dataset. This performance is not indicative of the downstream MPC performance of a target building that is not part of the validation set. 

\begin{table}[!b]
\centering
\footnotesize
\caption{Hyperparameter selection for all architectures.}
\label{tab_gm_hyperparams}
\begin{tabular}{l | c c c}
\toprule
\textbf{Parameter} & \textbf{MLP} & \textbf{LSTM} & \textbf{Transformer} \\
\hline
Hidden size       & 125  & 192  & 192  \\
\# Layers         & 3  & 3    & 3    \\
Lookback         & 96  & 96    & 96    \\
Batch size        & 64  & 96   & 128  \\
Learning rate     & .001  & .001 & .0004 \\
\# Attention heads & -  & -    & 4    \\
\bottomrule
\end{tabular}
\end{table}

\section{Cost Function Weights}
\label{app_weighting}

Table~\ref{tab_qu_cost} assesses the robustness of our method to the choice of cost weightings (Eq.~\ref{eq_mpc_obj}), comparing its dependence on the control performance across the 32 target buildings. For this experiment, we use the linear MPC benchmark (see Section~\ref{sec_eval}) and the generalized model pretrained on excitation-based data (MLP pretrained on ramp excitation), which are also used throughout the experiments in Section~\ref{sec_bench_ex}.

Table~\ref{tab_qu_cost} shows that the MLP achieves the lowest cost across all configurations, demonstrating consistent performance regardless of the chosen weighting combinations. In terms of normalized cost, the lowest values occur at $w_0=100$ and $w_1=1$, which is, for example, also used in \cite{BUNNING2022118491}. The linear model, in contrast, exhibits a large upper quartile (10401146) at this weighting, resulting in a normalized average cost of 87785. The reason for the large upper quartile is that a larger $w_0$ relative to the constant $w_2$ shifts the controller's priority toward correcting temperature tracking error rather than limiting control-input variation. This allows the controller to make larger adjustments to the control input between consecutive time steps. As a result, this may produce oscillations that incur additional costs over the control horizon due to the linear model's less accurate dynamic predictions (see Appendix~\ref{sec_pred_lin}). This contrast highlights that the generalized modeling approach is a more reliable choice independent of the specific weighting.

For this study, however, we use $w_0=w_1=1$ to enable a fair comparison with the linear benchmark. Weighting the heating term $w_1$ more heavily than the comfort term $w_0$ is uncommon in practice, as it would undermine the primary purpose of the heating system, which is to maintain thermal comfort \cite{stoffel_real-life_2024, STOFFEL2023112709, Drgona.2020}.

\begin{table}[t]
\centering
\caption{Median total and normalized cost with interquartile range for different cost function weights $w_0$ and $w_1$ (Eq.~\ref{eq_mpc_obj}) across target buildings. We keep $w_2$ constant, so that adjustments in $w_0$ and $w_1$ directly affect relative weighting in Eq.~\ref{eq_mpc_obj} against $w_2$, which also reduces the number of runs required. Each weight trial is executed over the one-month test window. Bold values indicate the lower median cost.}
\label{tab_qu_cost}
\resizebox{\columnwidth}{!}{%
\begin{tabular}{c rr rl rl}
\toprule
& $w_0$ & $w_1$ & \multicolumn{2}{c}{MLP - Ramp} & \multicolumn{2}{c}{Linear model} \\
\midrule
\multirow{5}{*}{\rotatebox{90}{Total cost}}
& 1   & 1   & \textbf{650}   & $\pm$ [592, 713]        & 702    & $\pm$ [619, 764]        \\
& 10  & 1   & \textbf{824}   & $\pm$ [732, 940]        & 863    & $\pm$ [780, 830441]     \\
& 100 & 1   & \textbf{3848}  & $\pm$ [2266, 5502]      & 6610   & $\pm$ [1566, 10401146]  \\
& 1   & 10  & \textbf{6825}  & $\pm$ [5897, 7443]      & 7443   & $\pm$ [6375, 9109]      \\
& 1   & 100 & \textbf{68003} & $\pm$ [57320, 76029]    & 130694 & $\pm$ [108438, 163169]  \\
\midrule
\multirow{5}{*}{\rotatebox{90}{Normalized}}
& 1   & 1   & \textbf{653.6} & $\pm$ [583.9, 703.1]    & 702.4   & $\pm$ [618.9, 763.5]     \\
& 10  & 1   & \textbf{82.4}  & $\pm$ [73.2, 94.0]      & 86.3    & $\pm$ [78.0, 83044.2]    \\
& 100 & 1   & \textbf{38.5}  & $\pm$ [22.7, 55.0]      & 66.1    & $\pm$ [15.7, 104011.5]   \\
& 1   & 10  & \textbf{682.5} & $\pm$ [589.7, 744.3]    & 744.3   & $\pm$ [637.5, 910.9]     \\
& 1   & 100 & \textbf{680.0} & $\pm$ [573.2, 760.3]    & 1306.9  & $\pm$ [1084.4, 1631.7]   \\
\bottomrule
\end{tabular}%
}
\end{table}

\section{Building-Level Control Performance}
\label{app_build_level}

Table~\ref{tab_building_cost} reports the total MPC cost per building for the four methods. The MLP pretrained on ramp excitation achieves the lowest cost for 25 of the 32 buildings; the online linear model performs best for the remaining 7.

\begin{table*}[t]
\centering
\caption{Per-building total MPC cost for each method with building properties displayed for the targets. Bold values indicate the best control performance.}
\label{tab_building_cost}
\begin{tabular}{lllllc|cccc}
\toprule
$U-value_{\text{wall}}$ & $c_{\text{wall}}$ & $A_{\text{ground}}$ & Weather & $f_{\text{win}}$ & Night set- & MLP & LSTM & Linear & PI \\
{[}W/(m\textsuperscript{2}K){]} & [kJ/(m\textsuperscript{2}K)] & [m\textsuperscript{2}] & & & back [K] & Ramp & Standard & model & controller \\
\midrule
0.25 & 40 & 70 & Amsterdam & 0.19 & 3.5 & \textbf{269.8} & 448.6 & 310.7 & 486.5 \\
0.25 & 150 & 100 & Amsterdam & 0.19 & 0.0 & \textbf{282.1} & 400.7 & 295.1 & 463.1 \\
0.25 & 150 & 100 & Bratislava & 0.16 & 3.0 & \textbf{582.7} & 1652.7 & 670.4 & 848.9 \\
0.25 & 150 & 100 & Munich & 0.16 & 0.0 & \textbf{712.0} & 2763.4 & 846.7 & 998.0 \\
0.25 & 150 & 70 & Munich & 0.19 & 3.0 & \textbf{664.2} & 3531.5 & 759.3 & 940.1 \\
0.25 & 280 & 100 & Amsterdam & 0.19 & 3.5 & \textbf{283.5} & 477.6 & 297.9 & 456.4 \\
0.25 & 280 & 100 & Bratislava & 0.19 & 4.0 & \textbf{604.5} & 1107.4 & 620.2 & 755.3 \\
0.25 & 280 & 100 & Munich & 0.19 & 3.0 & \textbf{643.6} & 2666.0 & 735.4 & 898.8 \\
0.25 & 280 & 70 & Amsterdam & 0.19 & 2.0 & \textbf{288.0} & 453.4 & 321.6 & 465.1 \\
0.25 & 280 & 70 & Munich & 0.16 & 1.0 & \textbf{630.1} & 1769.0 & 735.7 & 895.9 \\
0.25 & 280 & 70 & Munich & 0.19 & 3.0 & \textbf{595.6} & 2309.2 & 776.3 & 968.5 \\
0.55 & 40 & 100 & Bratislava & 0.19 & 2.0 & \textbf{648.6} & 3359.0 & 649.2 & 727.5 \\
0.55 & 40 & 70 & Munich & 0.16 & 2.0 & \textbf{754.3} & 5948.3 & 808.9 & 859.8 \\
0.55 & 280 & 100 & Munich & 0.19 & 0.0 & \textbf{668.7} & 3975.2 & 727.0 & 779.9 \\
0.55 & 280 & 70 & Munich & 0.19 & 0.0 & \textbf{651.5} & 5508.9 & 873.3 & 891.4 \\
0.85 & 40 & 100 & Bratislava & 0.16 & 0.0 & \textbf{677.5} & 7466.2 & 799.4 & 909.7 \\
0.85 & 40 & 100 & Bratislava & 0.19 & 3.0 & 673.5 & 4973.4 & \textbf{655.4} & 960.9 \\
0.85 & 40 & 100 & Munich & 0.19 & 3.0 & 846.0 & 8773.8 & \textbf{777.9} & 1058.0 \\
0.85 & 40 & 70 & Bratislava & 0.19 & 1.0 & 717.2 & 4259.4 & \textbf{630.5} & 964.1 \\
0.85 & 150 & 100 & Bratislava & 0.16 & 0.0 & \textbf{689.1} & 5853.5 & 741.3 & 862.5 \\
0.85 & 150 & 100 & Munich & 0.19 & 0.0 & 851.2 & 6102.1 & \textbf{746.9} & 1043.6 \\
0.85 & 150 & 70 & Bratislava & 0.19 & 0.0 & \textbf{626.0} & 4451.9 & 697.8 & 880.9 \\
0.85 & 150 & 70 & Munich & 0.19 & 2.5 & 749.4 & 5931.0 & \textbf{706.6} & 999.4 \\
0.85 & 280 & 70 & Amsterdam & 0.16 & 0.0 & \textbf{317.3} & 1060.0 & 350.1 & 452.4 \\
0.85 & 280 & 70 & Bratislava & 0.16 & 0.0 & \textbf{613.1} & 5341.9 & 698.2 & 835.6 \\
1.15 & 40 & 100 & Munich & 0.19 & 4.0 & 860.0 & 17462.3 & \textbf{814.1} & 1906.2 \\
1.15 & 150 & 100 & Bratislava & 0.16 & 1.0 & \textbf{634.4} & 7111.8 & 686.9 & 1584.1 \\
1.15 & 150 & 70 & Munich & 0.16 & 0.0 & \textbf{789.8} & 13125.4 & 879.0 & 1761.5 \\
1.15 & 280 & 100 & Amsterdam & 0.16 & 0.0 & \textbf{346.6} & 1847.4 & 398.1 & 853.1 \\
1.15 & 280 & 100 & Bratislava & 0.16 & 1.0 & \textbf{705.0} & 10218.8 & 732.8 & 1313.3 \\
1.15 & 280 & 70 & Amsterdam & 0.16 & 0.0 & \textbf{306.0} & 1716.9 & 398.6 & 767.6 \\
1.15 & 280 & 70 & Bratislava & 0.19 & 1.5 & 748.6 & 5024.3 & \textbf{615.0} & 2242.8 \\
\bottomrule
\end{tabular}
\end{table*}

\newpage
\section{Prediction Performance of Linear Model}
\label{sec_pred_lin}

\begin{figure}
	\centering
	\includegraphics[width=0.88\columnwidth]{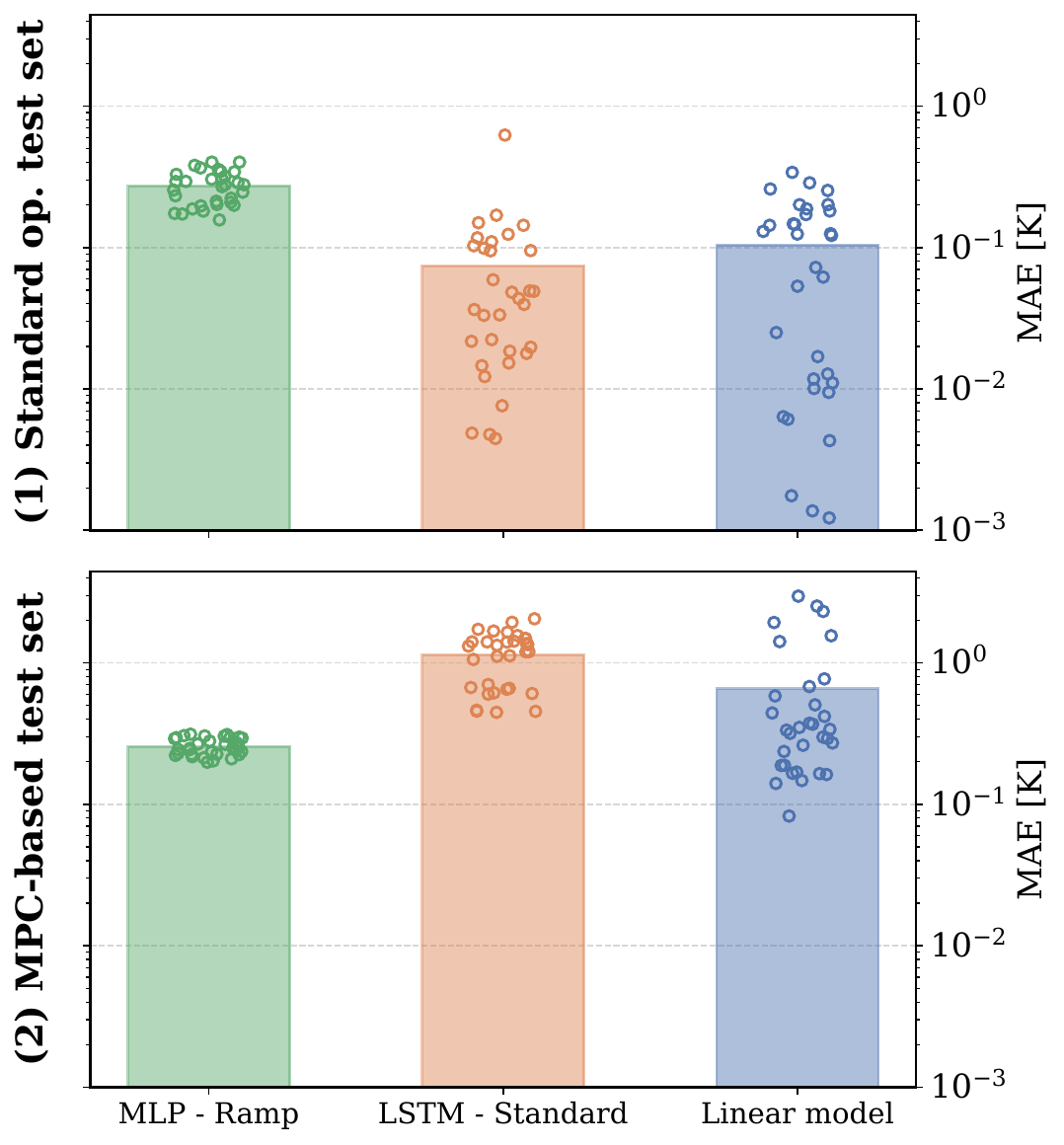}
	\caption{Prediction accuracy (MAE in log scale) of the MLP pretrained on ramp excitation, the LSTM pretrained on standard operational data, and the linear model (first online iteration), on the standard operation test set (1) and the MPC-based test set (2) target data. Bars show the mean across target buildings; markers show individual buildings.}
\label{ex_box_linear_pred}
\end{figure}

Figure~\ref{ex_box_linear_pred} compares the prediction accuracy of the linear model against the MLP pretrained on ramp excitation and the LSTM pretrained on standard operational data, evaluated under both the standard operation test set (1) and the MPC-based test set (2), as discussed in Section~\ref{sec_bench_ex}. For the linear model, we use the model obtained after the first online training cycle, meaning it used only December data. Testing is performed on January data. The linear model achieves a mean MAE of 0.10~K, close to that of the LSTM trained on standard operational data (MAE of 0.08~K) under the standard operation test set (1), though with higher variability across buildings. Under the MPC-based test set (2), the linear model's mean prediction error increases to 0.66~K, and its spread across buildings widens, although most buildings remain below an MAE of 1~K. This results in a prediction gap of 0.56~K between the two test sets, smaller than that of the LSTM trained on standard operational data (1.07~K), but larger than that of the MLP trained on ramp excitation (0.016~K). 

\newpage
\section{Ten-Week Control Evaluation}
\label{app_3M_control}

\begin{figure}
	\centering
	\includegraphics[width=0.5\textwidth]{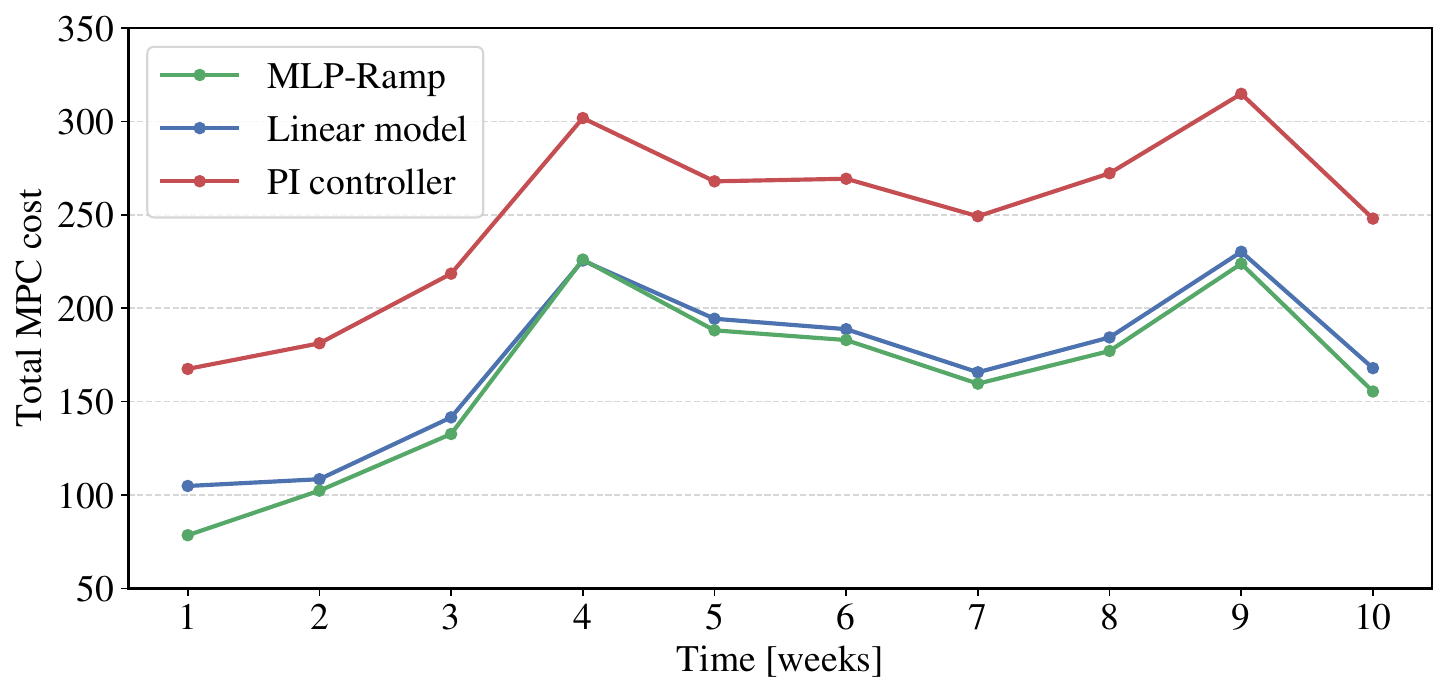}
	\caption{Total MPC cost per week over the ten-week evaluation period averaged weekly over all target buildings.}
	\label{ex_cost_over_time}
\end{figure}

Figure~\ref{ex_cost_over_time} shows the mean MPC cost over the ten-week evaluation period per week across all target buildings. All three controllers follow a broadly similar temporal pattern, with cost increasing slightly from week 1 to a peak around weeks 4--9, before decreasing again toward the end of the period. This temporal trend is primarily driven by weather patterns. Across all periods, the MLP pretrained on ramp excitation achieves the best performance, closely followed by the linear model.

\newpage
\balance
\bibliographystyle{elsarticle-num}

\bibliography{references.bib}


\end{document}